\documentclass[12pt]{article}
\usepackage{amsmath,amsfonts,amssymb,amsthm}
\usepackage{array}
\usepackage{cite}
\usepackage{enumerate}
\usepackage{hyperref}
\hypersetup{bookmarksopen,bookmarksnumbered,pdfstartview=FitH,colorlinks,citecolor=blue,unicode}
\usepackage{indentfirst}
\usepackage[low-sup]{subdepth}
\usepackage{relsize}
\usepackage{empheq} 	
\usepackage{tensor}
\usepackage{authblk}
\newcommand\numberthis{\addtocounter{equation}{1}\tag{\theequation}}
\newcommand\nachoo[1]{\mathring{\mathlarger{\rhd}}_{#1}}
\newcommand\nacho[1]{\mathlarger{\rhd}_{#1}}
\newcommand\uppernacho[1]{\mathlarger{\rhd}^{#1}}
\newcommand\uppernachoo[1]{\mathring{\mathlarger{\rhd}}^{#1}}
\newcommand\f[2]{\tensor{f}{_{#1}^{#2}}}
\newcommand\g[2]{\tensor{g}{_{#1}^{#2}}}
\newcommand\ginv[2]{(g^{-1})\tensor{}{_{#1}^{#2}}}
\newcommand\e[2]{\tensor{e}{_{#1}^{#2}}}
\newcommand\einv[2]{(e^{-1})\tensor{}{_{#1}^{#2}}}
\title{\vskip-2in\hbox to\textwidth{\hfill \normalsize YITP-SB-16-26}\vskip1.8in Gauging Unbroken Symmetries in {\bf F}-theory}
\author{Chia-Yi Ju\thanks{\href{mailto:cju@insti.physics.sunysb.edu}{cju@insti.physics.sunysb.edu}} and Warren Siegel\thanks{\href{mailto:siegel@insti.physics.sunysb.edu}{siegel@insti.physics.sunysb.edu}}}
\affil{\it C. N. Yang Institute for Theoretical Physics\\
\it State University of New York, Stony Brook, NY 11794-3840}
\date{}

\begin{document}
	\numberwithin{equation}{section}
	\maketitle
	\begin{abstract}
{\bf F}-theory attempts to include all {\bf U}-dualities manifestly. Unlike its {\bf T}-dual manifest partner, which is based on string current algebra, {\bf F}-theory is based on higher dimensional brane current algebra. Like the {\bf T}-dual manifest theory, which has $O(D-1,1)^2$ unbroken symmetry, the {\bf F}-theory vacuum also enjoys certain symmetries (``$H$"). One of its important and exotic properties is that worldvolume indices are also spacetime indices. This makes the global brane current algebra incompatible with $H$ symmetry currents. 
The solution is to introduce worldvolume covariant derivatives, which depend on the $H$ coordinates even in a ``flat" background.
We will also give as an explicit example the 5-brane case.
		
	\end{abstract}
	
	\section{Introduction}
	
		It is well known that many string theories (``{\bf S}-theories") are related by different dualities \cite{Giveon:1994fu,Seiberg:1994pq,Hull:1994ys}. On the one hand, this led to the idea of {\bf M}-theory \cite{Witten:1995ex,Horava:1995qa} (a theory that is one dimension higher than {\bf S}-theory) that links all different types of {\bf S}-theories to each other by dualities. It's also known that {\bf M}-theory (or {\bf S}-theory) would further imply {\bf U}-duality \cite{Hull:1994ys,Cremmer:1997ct,Obers:1998fb}, conjectured to be a discrete subgroup of $E_{n(n)}$ ($n$ is the dimension of {\bf M}-theory), which is the most general duality (including {\bf S}-duality and {\bf T}-duality) of {\bf S}-theory. However, {\bf M}-theory cannot give us every type of {\bf S}-theory without using duality \cite{Vafa:1996xn}, i.e. the theory does not manifest all dualities. On the other hand, {\bf T}-theory \cite{Siegel:1993xq,Siegel:1993th,Siegel:1993bj} (sometimes called ``Double Field Theory" \cite{Hull:2009mi}) has {\bf T}-duality manifest by including the ``{\bf T}-dual'' coordinates  (coordinates are doubled).
		
		{\bf F}-theory \cite{Hatsuda:2012uk,Hatsuda:2013dya,Linch:2015lwa,Linch:2015fya,Linch:2015fca,Linch:2015qva} is meant to take advantage of both {\bf M}-theory and {\bf T}-theory -- having {\bf U}-dualities manifest by including all the {\bf U}-dual coordinates. To accomplish that, we have to use branes rather than strings. The low energy limit of that is a theory with $G = E_{n(n)}$ symmetry. However, some of the symmetries are also symmetries of the vacuum, which form the group $H$. Therefore, we can treat $H$ as a gauge group. Since {\bf F}-theory includes both {\bf M}-theory and {\bf T}-theory, $H$ should at least include both the symmetries of the vacuum in {\bf M}-theory and {\bf T}-theory.
		 
Although the currents of the theory have been found \cite{Linch:2015fca}, it is usually easier to work with by having $H$ symmetry manifest instead of gauging them to zero directly. 
We therefore use the group element $g$ to make $H$ symmetry local even in ``flat'' spacetime.
The group coordinates are then included with the other ``spacetime" coordinates as worldvolume fields.
However, we found that the global current algebra is not compatible with $H$ symmetry currents. 
This requires that the derivatives of $\delta$-functions that appear in the Schwinger terms be covariantized with $g$.
These derivatives were found previously to need covariantization in nontrivial backgrounds, but when gauging $H$ even ``flat" spacetime has a vielbein that is not constant.
		
		In this paper, we start by showing the global brane current algebra does not go along with $H$ symmetry currents. Then we modify the theory and give a very general construction for arbitrary finite dimensional current algebras and check its consistency with Jacobi identities.  We will give an explicit example for the 5-brane case, where the $H$ group is $Spin(3,2)$.

	\section{Notations}
	
		Instead of explaining our notations all over the paper, some common notations are defined in this section so that the readers don't have to hunt for them.
		
		\begin{enumerate}[i)]
			\item $\displaystyle f(1) \equiv f(\sigma_1)$, where $\sigma$ is worldvolume coordinates, $\displaystyle f(1 - 2) \equiv f(\sigma_1 - \sigma_2)$, $\displaystyle f((1) + (2)) \equiv f(1) + f(2)$, and, similarly, $\displaystyle f((1) - (2)) \equiv f(1) - f(2)$.
			\item Worldvolume vector indices are denoted as $q$, $r$, $\cdots$; spacetime spinor indices are $\alpha$, $\beta$, $\cdots$; superspace indices (which include $\lbrace D, P, \Omega \rbrace$) are $M$, $N$, $O$, $\cdots$; covariantized superspace indices are $A$, $B$, $C$, $\cdots$ (also include $\lbrace D, P, \Omega \rbrace$); group coordinate indices are denoted as $I$, $J$, $K$, $\cdots$; the covariantized index for $H$ group is $S$, and the full set of covariantized superspace indices, including all ``$A$'' indices, $S$, and $\Sigma$, are $\mathcal{A}$, $\mathcal{B}$, $\mathcal{C}$, $\cdots$.
			\item $\displaystyle \nachoo{M}(\sigma) = $ worldvolume current:
				e.g. $\displaystyle \nachoo{D}(\sigma) = \nachoo{\alpha}$, $\displaystyle \nachoo{\Omega}(\sigma) = \uppernachoo{\alpha r}$.
			\item $\displaystyle \eta_{MNr}$ is the generalized constant metric, $\f{MN}{O}$ is the structure constants.
			\item $\alpha^I (\sigma)$ is the coordinates of $H$ group (a function of the worldvolume).
			\item $\displaystyle \e{S}{I}(\sigma)$ is the vielbein that converts functional derivatives $\displaystyle \left(\frac{\delta}{\delta \alpha^I (\sigma)}\right)$ into symmetry generators $\displaystyle \left(\nacho{S}(\sigma)\right)$.
			\item $\displaystyle \partial^r = \frac{\partial}{\partial \sigma_r}$, a worldvolume coordinate derivative. Sometimes we have to specify which coordinate we act on:  Then we add an additional index $\displaystyle \partial^r_1 = \frac{\partial}{\partial \sigma_{1r}}$.
			\item $\displaystyle \nacho{A}(\sigma) = $ covariantized worldvolume current.
			\item $\displaystyle \nacho{\mathcal{A}}(\sigma) = $ the full set of covariant worldvolume currents.
			\item $\g{A}{M}(\sigma)$ is a worldvolume field and is an element of $H$ group.
			\item Parenthesis $[~)$ in $\displaystyle f_{[m|n|o)}$ is the graded (anti)symmetrization, i.e. sum of index permutation (with a minus sign if not interchanging two spinor indices) in the parenthesis but not the ones in between the two vertical lines, $|~|$.
		\end{enumerate}
				
	\section{General Construction\label{GC}}
		\subsection{Problem With Naive Approach}
			
			We start this subsection with an observation, and then work the way to the general case, showing why the naive current algebra is not compatible with global $H$ symmetry.
			
			As described in \cite{Linch:2015fya}, we know that in general the worldvolume current $\nachoo{M}$ obeys the following algebra:
			\begin{align*}
				\Big[ \nachoo{N}(1) , \nachoo{N}(2) \Big] = i \f{MN}{O} \nachoo{O} \delta(1-2) + 2 i \eta_{M N r} \partial^r_1 \delta(1 - 2).
			\end{align*}
			
			We could naively introduce additional $H$-group worldvolume currents $\nachoo{S}$ and force
			\begin{align*}
				\Big[ \nachoo{S}(1) , \nachoo{M}(2) \Big] = i\f{SM}{M'} \nachoo{M'} \delta(1 - 2).
			\end{align*}
			Then a symmetric part of Jacobi identities gives
			\begin{align}
				\tensor{f}{_{S M}^{M'}}\eta_{M' N r} + \f{S N}{N'}\eta_{M N' r} = 0. \label{contradiction}
			\end{align}
			However, this doesn't work because spacetime indices in F-theory are also worldvolume indices. If we want spacetime indices to transform under $H$ group, then worldvolume should transform as well. Since $\eta$ is a ``constant'' under group $G$-transformations, it should also be invariant under $H$-transformation. And we're led to the following identity:
			\begin{align*}
				\tensor{f}{_{S M}^{M'}}\eta_{M' N r} + \f{S N}{N'}\eta_{M N' r} + \f{Sr}{q}\eta_{M N q} = 0, 
			\end{align*}
			which contradicts equation (\ref{contradiction}).
		
To find a solution, it is useful to go back to the general construction of symmetry generators of $H$ group. The generalized symmetry generator method from particle to brane is listed in the table below:
\begin{align*}
\begin{array}{|l|l|}
\hline
\text{Particle} & \text{Brane} \\
\hline
\hbox{\vrule height14pt depth7pt width0pt}(dg)g^{-1} = dx^i \einv{i}{s} G_{s} & (\delta g)g^{-1} = \delta\alpha^I \einv{I}{S} G_{S}\\
\hbox{\vrule height14pt depth7pt width0pt}\nacho{s} = i \e{s}{i}\frac{\partial}{\partial x^i} & \nachoo{S} = i \e{S}{I}\frac{\delta}{\delta \alpha^I}\\
\hbox{\vrule height14pt depth7pt width0pt} & \uppernachoo{Sr} = \left( \partial^r \alpha^I \right) \einv{I}{S}\\
\hline
\end{array}
\end{align*}
			Here $G$ is the symmetry generator of the group. The last brane current is the ``dual'' current of $\nacho{S}$, which does not have a particle analog. The relations between structure constants and generalized metric are listed as follows:
			\begin{align*}
				&& \Big[ \nachoo{S_1}(1) , \nachoo{S_2}(2) \Big\rbrace & = -\e{[S_1|}{I}\left(\frac{\delta}{\delta \alpha^I} \e{|S_2)}{J}\right) \einv{J}{O} \e{O}{L} \frac{\delta}{\delta \alpha^L}\delta(1-2)\\
				&&& = i \e{[S_1|}{I}\left(\frac{\delta}{\delta \alpha^I} \e{|S_2)}{J}\right) \einv{J}{O} \nachoo{O}\delta(1-2)\\
				&&& = i \f{MN}{O}\nachoo{O}\delta(1-2),\\
				&& \Big[ \nachoo{S_1}(1) , \uppernachoo{S_2 r}(2) \Big\rbrace & = i \e{P}{I}\left(\frac{\delta}{\delta \alpha^I} \e{S_1}{J}\right) \einv{J}{O} \e{O}{S_2} \left( \partial^r \alpha^L \right) \einv{L}{P}\delta(1-2)\\
				&&& + i \delta_{S_1}^{S_2}\partial^r_2\delta(1-2)\\
				&&& = i \f{PM}{N}\uppernachoo{Pr}\delta(1-2) + i \tensor{\eta}{_{S_1}^{S_2 r}_q}\partial^q_2 \delta(1-2),\\
				&& \Big[ \uppernachoo{S_1 r}(1) , \uppernachoo{S_2 q}(2) \Big\rbrace & = 0.
			\end{align*}
			It's useful to inspect the simplest case with metric only (we neglect $f$ term):
			\begin{align*}
				\Big[ \nachoo{S_1}(1) , \uppernachoo{S_2r}(2) \Big\rbrace \sim i \tensor{\eta}{_{S_1}^{S_2 r}_q}\partial^q_2 \delta(1-2).
			\end{align*}
			In {\bf F}-theory, again, worldvolume indices are also spacetime indices. They both have to transform the same way under $H$ group. However, by construction, $\partial^r$ is not a function of $\alpha$, therefore, both $r$ and $q$ don't transform under $\nachoo{S}$. We are led to an impasse.
			
			For the rest of the paper we will denote $\nachoo{\Sigma}$ instead of $\uppernacho{S r}$ for simplicity. 
		\subsection{Solution}
			To include the group $H$ in the theory, we introduce a set of $H$ group coordinates as worldvolume fields $\alpha^I(\sigma)$, and the corresponding group elements $g(\alpha(\sigma))) \in H$, and their inverses. By definition, they obey the following commutation relation:
			\begin{align*}
				& \Big[ \nacho{S}(1) , \g{A}{M}(2) \Big] = i \f{S A}{B}\g{B}{M} \delta(1 - 2),\\
				& \Big[ \nacho{S}(1) , \ginv{M}{A}(2) \Big] = i \ginv{M}{B} \f{BS}{A} \delta(1-2),
			\end{align*}
			where $\nacho{S}$ is the symmetry generator of group $H$. We also define new sets of currents by multiplying the old ones with $g's$, i.e.
			\begin{align*}
				\nacho{A} \equiv \g{A}{M}\nachoo{M},
			\end{align*}
			so that all the indices transform under $H$-group as well.
			
			We should point out that since we have introduced a current $\nacho{S}$, we should also introduce its dual current $\nachoo{\Sigma}$:
			\begin{align*}
				\nachoo{\Sigma} \equiv \uppernachoo{Sr} = \left(\partial^{r} \alpha^I\right)\einv{I}{S}.
			\end{align*}
			
			It can be shown that the original metrics are unaffected if the worldvolume derivative is also multiplied by $g$:
			\begin{align*}
				& \Big[ \nachoo{M}(1) , \nachoo{N}(2) \Big\rbrace = i \f{MN}{O}\nachoo{O} \delta(1-2) + 2 i \eta_{MNr}\partial^r_1 \delta(1-2)\numberthis\label{compare}\\
				\Rightarrow & \Big[ \nacho{A}(1) , \nacho{B}(2) \Big\rbrace = \Big[ \g{A}{M}\nachoo{M}(1) , \g{B}{N} \nachoo{N}(2) \Big\rbrace\\
				= &\ i \g{A}{M} \g{B}{N} \f{MN}{O}\nachoo{O} \delta(1-2) + 2 i \g{A}{M} (1) \g{B}{N} (2) \eta_{MNr}\partial^r_1 \delta(1-2)\\
				= &\ i \g{A}{M} \g{B}{N} \f{MN}{O}\ginv{O}{C}\nacho{C} \delta(1-2) + 2 i \g{A}{M} (1) \g{B}{N} (2) \eta_{MNr}\partial^r_1 \delta(1-2)\\
				= &\ i \g{A}{M} \g{B}{N} \f{MN}{O}\ginv{O}{C}\nacho{C} \delta(1-2) + i \g{[A|}{M} \partial^r \g{|B)}{N} \eta_{MNr} \delta(1-2)\\
				& + i \left(\g{A}{M} \g{B}{N} \eta_{MNr} \right)((1) + (2)) \partial^r_1 \delta(1-2)\\
				= &\ i \g{A}{M} \g{B}{N} \f{MN}{O}\ginv{O}{C}\nacho{C} \delta(1-2) + i \g{[A|}{M} \partial^r \g{|B)}{N} \eta_{MNr} \delta(1-2)\\
				& + i \left(\g{A}{M} \g{B}{N} \g{a}{p} \ginv{r}{a} \eta_{MNp} \right)((1) + (2)) \partial^r_1 \delta(1-2)\\
				= &\ i \f{AB}{C}\nacho{C} \delta(1-2) + i \g{[A|}{M} \partial^r \g{|B)}{N} \eta_{MNr} \delta(1-2)\\
				& + i \eta_{ABa}\ginv{r}{a}((1) + (2)) \partial^r_1 \delta(1-2)\\
				= &\ i \f{AB}{C}\nacho{C} \delta(1-2) + i \g{[A|}{M} \partial^r \g{|B)}{N} \eta_{MNr} \delta(1-2)\\
				& + i \eta_{ABa}\uppernacho{a} ((1)-(2)) \delta(1-2),
			\end{align*}
			where $\f{AB}{C} = \g{A}{M} \g{B}{N} \f{MN}{O}\ginv{O}{C}$, $\eta_{ABa} = \g{A}{M} \g{B}{N} \g{a}{r} \eta_{MNr}$, and 
$$\uppernacho{a} = \ginv{r}{a} \partial^r$$
Since both $f$'s and $\eta$'s are invariant under $H$-group, $\f{AB}{C}$ and $\eta_{ABa}$ are numerically equal to $\f{MN}{O}$ and   $\eta_{MNr}$ respectively.  The term $\g{[A|}{M} \partial^r \g{|B)}{N} \eta_{MNr} \delta(1-2)$ is in fact a torsion term:
			\begin{align*}
				\g{[A|}{M} \partial^r \g{|B)}{N} \eta_{MNr} & = \g{[A|}{M} \partial^r \g{|B)}{O} \ginv{O}{C} \g{C}{N}\eta_{MNr}\\
				& = \g{[A|}{M} \left(\partial^r \alpha^I\right)\left(\frac{\delta}{\delta \alpha^I}\g{|B)}{O}\right) \ginv{O}{C} \g{C}{N}\eta_{MNr}\\
				& = \g{[A|}{M} \left(\partial^r \alpha^{I}\right)\einv{I}{S}\left(G_{S}\right)\tensor{ }{_{|B)}^{C}}\g{C}{N}\eta_{MNr}\\
				& = - \left(\uppernacho{a} \alpha^{I}\right)\einv{I}{S}\f{S[A|}{C}\eta_{C|B)a}\\
				& = \uppernacho{Sa}\f{S[A|}{C}\eta_{C|B)a}\\
				& = \nacho{\Sigma}\f{AB}{\Sigma},
			\end{align*}
			where
			\begin{align*}
				\nacho{\Sigma} = \uppernacho{Sa} \equiv \ginv{r}{a}\uppernachoo{Sr}.
			\end{align*}
			The third equality comes from $(\delta g)g^{-1} = (\delta \alpha) e^{-1} G$. For the fourth equality we use the fact that $\left(G_a\right)\tensor{}{_b^c} = \f{ab}{c}$ in adjoint representation. The $\displaystyle \left(\uppernacho{a} \alpha^{I}\right)\einv{I}{S}$ in the fourth line is the covariant ``dual'' of $\nacho{S}$ ($\nacho{\Sigma} = \uppernacho{\widetilde{S}a}$). Using that fact that $\eta_{S\Sigma a} = \tensor{\eta}{_{S}^{\widetilde{Sb}}_a} = - \frac{1}{2}\delta_{S}^{\widetilde{S}}\delta_a^b$ (the $-\frac{1}{2}$ comes from the definition of equation (\ref{compare})), we get
			\begin{align}
				\f{AB}{\Sigma}\eta_{\Sigma S a} = \frac{1}{2}\f{S[A|}{C}\eta_{C|B)a}. \label{fssigma}
			\end{align}
			We close this section by calculating the commutation relation between $\nacho{S}$ and $\nacho{\Sigma}$:
			\begin{align*}
				& \Big[ \nachoo{S}(1) , \nachoo{\Sigma}(2) \Big] = \Big[ \nachoo{S_1}(1) , \uppernachoo{S_2a}(2) \Big]\\
				= &\ \Big[ i \e{S_1}{I} \frac{\delta}{\delta \alpha^{I}}(1) , \ginv{r}{a}\left(\partial^r \alpha^{J} \right) \e{J}{S_2} \Big]\\
				= &\ i \f{bS_1}{a}\ginv{r}{b}\left(\partial^r \alpha^{I} \right) \e{I}{S_2} \delta(1-2) + i \left( \e{S_1}{I} \frac{\partial}{\partial \alpha^{I}} \einv{J}{S_2}\right) \ginv{r}{a} \left(\partial^r \alpha^{J} \right)\delta(1-2)\\
				& + i \e{S_1}{I}(1)\ginv{r}{a}(2)\einv{I}{S_2}(2)\partial^r_2 \delta(1-2)\\
				= &\ i \f{bS_1}{a}\ginv{r}{b}\left(\partial^r \alpha^{I} \right) \e{I}{S_2} \delta(1-2) + i \left( \e{S_1}{I} \frac{\partial}{\partial \alpha^{I}} \einv{J}{S_2}\right) \ginv{r}{a}\left(\partial^r \alpha^{J} \right)\delta(1-2)\\
				& + i \left(\partial^r \e{S_1}{I}\right)\ginv{r}{a}\einv{I}{S_2}\delta(1-2) + i \left(\e{S_1}{I}\ginv{r}{a}\einv{I}{S_2}\right)(2)\partial^r_2 \delta(1-2)\\
				= &\ i \f{bS_1}{a}\ginv{r}{b}\left(\partial^r \alpha^{I} \right) \e{I}{S_2} \delta(1-2) \\
				& - i \e{S_1}{I} \einv{J}{S_3}\left(\frac{\partial}{\partial \alpha^{I}} \e{S_3}{K}\right)\einv{K}{S_2} \ginv{r}{a}\left(\partial^r \alpha^{J} \right)\delta(1-2)\\
				& + i \e{S_3}{I}\einv{J}{S_3}\left(\frac{\partial}{\partial \alpha^{I}} \e{S_1}{K}\right)\einv{K}{S_2} \ginv{r}{a}\left(\partial^r \alpha^{J} \right)\delta(1-2) + i \delta_{S_1}^{S_2}\uppernacho{a}(2) \delta(1-2)\\
				= &\ i \f{bS_1}{a}\ginv{r}{b}\left(\partial^r \alpha^{I} \right) \e{I}{S_2} \delta(1-2) + i \f{S_3S_1}{S_2} \einv{J}{S_3} \left(\partial^r \alpha^{J} \right)\delta(1-2) \\
				& + i \delta_{S_1}^{S_2}\uppernacho{a}(2) \delta(1-2)\\
				= &\ i \f{bS_1}{a} \uppernacho{S_2b} \delta(1-2) + i \f{S_3S_1}{S_2} \uppernacho{S_3a}\delta(1-2) + i \delta_{S_1}^{S_2}\uppernacho{a}(2) \delta(1-2)\\
				= &\ i \f{S \Sigma}{\Sigma'} \nacho{\Sigma'} \delta(1-2) + i \eta_{S \Sigma a}\uppernacho{a}(2) \delta(1-2).
			\end{align*}
	\section{Jacobi Identity}
		In the last section, we have constructed a very general mechanism to find all the currents. We will now check the Jacobi identity between currents and check if they are consistent with the method described above. Here we mention some results in \cite{Linch:2015fca}:  The worldvolume currents are $\lbrace \nachoo{D}, ~ \nachoo{P}, ~ \nachoo{\Omega} \rbrace$ with the following nonvanishing commutation relations:
		\begin{align*}
			& \Big\lbrace \nachoo{D_1}(1) , \nachoo{D_2}(2) \Big\rbrace = i \f{D_1 D_2}{P}\nachoo{P} \delta(1-2),\\
			& \Big[ \nachoo{D}(1) , \nachoo{P}(2) \Big] = i \f{D_1 P}{\Omega}\nachoo{\Omega}\delta(1-2),\\
			& \Big[ \nachoo{P_1}(1) , \nachoo{P_2}(2) \Big] = 2 i \eta_{P_1 P_2 r} \partial^r_1 \delta(1-2)\\
			& \Big\lbrace \nachoo{D}(1) , \nachoo{\Omega}(2) \Big\rbrace = 2 i \eta_{D \Omega r} \partial^r_2 \delta(1-2).
		\end{align*}
		We now generalize the above commutation relations using the method mentioned in Section \ref{GC} to include $H$ group; we get:
		\begin{align*}
			& \Big\lbrace \nacho{D_1}(1) , \nacho{D_2}(2) \Big\rbrace = i \f{D_1 D_2}{P}\nacho{P} \delta(1-2),\\
			& \Big[ \nacho{D}(1) , \nacho{P}(2) \Big] = i \f{D_1 P}{\Omega}\nacho{\Omega}\delta(1-2),\\
			& \Big[ \nacho{P_1}(1) , \nacho{P_2}(2) \Big] = i \f{P_1 P_2}{\Sigma}\nacho{\Sigma} + i \eta_{P_1 P_2 a} \uppernacho{a} ((1) - (2)) \delta(1-2)\\
			& \Big\lbrace \nacho{D}(1) , \nacho{\Omega}(2) \Big\rbrace = i \f{D \Omega}{\Sigma}\nacho{\Sigma} + i \eta_{D \Omega a} \uppernacho{a} ((1) - (2)) \delta(1-2),\\
			& \Big[ \nacho{S}(1) , \nacho{D}(2) \Big] = i \f{S D}{D'}\nacho{D'} \delta(1-2),\\
			& \Big[ \nacho{S}(1) , \nacho{P}(2) \Big] = i \f{S P}{P'}\nacho{P'} \delta(1-2),\\
			& \Big[ \nacho{S}(1) , \nacho{\Omega}(2) \Big] = i \f{S \Omega}{\Omega'}\nacho{\Omega'} \delta(1-2),\\
			& \Big[ \nacho{S}(1) , \nacho{S}(2) \Big] = i \f{S S}{S'}\nacho{S'} \delta(1-2),\\
			& \Big[ \nacho{S}(1) , \nacho{\Sigma}(2) \Big] = i \f{S \Sigma}{\Sigma'}\nacho{\Sigma'} \delta(1-2)+ 2 i \eta_{S \Sigma a} \uppernacho{a} (2) \delta(1-2).
		\end{align*}
		We can find all the relations between $f$'s and $\eta$'s by plugging in the above commutation relations into the Jacobi identities, which are listed in Appendix \ref{AppA}. Here we point out some of the interesting ones.
		The first example is combining equation (\ref{PSP}) and (\ref{SPP}),
		\begin{align*}
			\left\lbrace \begin{array}{l}
				0 = \tensor{f}{_{S P_1}^{P'}}\eta_{P' P_2 a} + \tensor{f}{_{S P_2}^{P'}}\eta_{P_1 P' a} + \tensor{f}{_{S a}^{b}}\eta_{P_1 P_2 b}\\
				0 = \tensor{f}{_{P_1 S}^{P'}}\eta_{P' P_2 a} + \tensor{f}{_{P_1 P_2}^{\Sigma'}}\eta_{S \Sigma' a} - \frac{1}{2}\tensor{f}{_{S a}^{b}}\eta_{P_1 P_2 b}
			\end{array} \right.
		\end{align*}
gives
		\begin{align*}
			\f{P_1 P_2}{\Sigma}\eta_{\Sigma S a} = \frac{1}{2}\f{S[P_1|}{P'}\eta_{P'|P_2]a},
		\end{align*}
		which is exactly the result in equation (\ref{fssigma}). It is worthwhile to point out that equation (\ref{DSOmega}) or (\ref{OmegaSD}) together with (\ref{SDOmega}) gives the same result as equation (\ref{fssigma}).
		
		And another interesting result is equation (\ref{SSigma}),
		\begin{align*}
			0 = \tensor{f}{_{S_1 S_2}^{S'}}\eta_{S' \Sigma a} + \tensor{f}{_{S_1 \Sigma}^{\Sigma'}}\eta_{S_2 \Sigma' a} + \tensor{f}{_{S_1 a}^{b}}\eta_{S_2 \Sigma b}.
		\end{align*}
		If we write the above equation explicit in $\widetilde{S}$ and $b$ (worldvolume) indices, we get:
		\begin{align*}
			0 & = \f{S_1 S_2}{S'}\tensor{\eta}{_{S'} ^{\widetilde{S} b} _{a}} + \tensor{f}{_{S_1} ^{\widetilde{S} b}_{\widetilde{S}' b'}}\tensor{\eta}{_{S_2} ^{\widetilde{S}' b'} _{a}} + \tensor{f}{_{S_1 a}^{a'}}\tensor{\eta}{_{S_2} ^{\widetilde{S} b} _{a'}}\\
			& = \f{S_1 S_2}{S'}\tensor{\eta}{_{S_2} ^{\widetilde{S} b} _{a}} + \f{\widetilde{S}' S_1}{\widetilde{S}}\tensor{\eta}{_{S_2} ^{\widetilde{S}' b} _{a}} + \tensor{f}{_{b' S_1}^{b}}\tensor{\eta}{_{S_2} ^{\widetilde{S} b'} _{a}} + \tensor{f}{_{S_1 a}^{a'}}\tensor{\eta}{_{S_2} ^{\widetilde{S} b} _{a'}}\\
			& = \f{S_1 S_2}{S'}\tensor{\eta}{_{S'} ^{\widetilde{S}}}\delta_{a}^{b} + \f{\widetilde{S}' S_1}{\widetilde{S}} \tensor{\eta}{_{S_2} ^{\widetilde{S}'}} \delta_{a}^{b} + \tensor{f}{_{b' S_1}^{b}}\tensor{\eta}{_{S_2} ^{\widetilde{S}}}\delta_{a}^{b'} + \tensor{f}{_{S_1 a}^{a'}}\tensor{\eta}{_{S_2} ^{\widetilde{S}}}\delta_{a'}^{b}\\
			& = \f{S_1 S_2}{S'}\tensor{\eta}{_{S'} ^{\widetilde{S}}}\delta_{a}^{b} + \f{\widetilde{S}' S_1}{\widetilde{S}}\tensor{\eta}{_{S'} ^{\widetilde{S}'}} \delta_{b}^{b}\\
			\Rightarrow 0 & = \f{S_1 S_2}{S'}\tensor{\eta}{_{S'} ^{\widetilde{S}}} + \f{\widetilde{S}' S_1}{\widetilde{S}}\tensor{\eta}{_{S_2} ^{\widetilde{S}'}},
		\end{align*}
		i.e. $\displaystyle \tensor{\eta}{_{S} ^{\widetilde{S}}}$ is $H$-invariant by itself.
	\section{Example: 5-brane}
		{\bf F}-theory on the 5-brane has been investigated quite intensively, e.g. \cite{Park:2013gaj,Linch:2015lwa,Linch:2015fya,Linch:2015fca,Linch:2015qva}. We go along with the trend and apply the above method to this case. It is shown in \cite{Linch:2015fca} that the bosonic sector lives in $SL(5)/SO(3,2)$. In order to be generalized to supersymmetry, rather than choosing $H = SO(3,2)$, we look for its double covering group $H = Spin(3,2) \cong Sp(4)$ and impose
		\begin{align*}
			\left\lbrace \nachoo{D_1} , \nachoo{D_2} \right\rbrace = i \f{D_1 D_2}{P}\nachoo{P}.
		\end{align*}
		The only invariant tensors that are symmetric in the two symmetric spinor indices are the Dirac $\gamma$-matrices with two antisymmetric vector indices, i.e. $\left(\gamma^{mn}\right)_{\alpha \beta}$. Hence, 
		\begin{align*}
			\left\lbrace \nachoo{D_1} (1) , \nachoo{D_2} (2) \right\rbrace = \left\lbrace \nachoo{\alpha_1} (1) , \nachoo{\alpha_2} (2) \right\rbrace =  i \left(\gamma^{mn}\right)_{\alpha_1 \alpha_2} \nachoo{mn} \delta(1-2),
		\end{align*}
		i.e. $\displaystyle \nachoo{P} = \nachoo{mn} = - \nachoo{nm}$.
		This then leads to
		\begin{align*}
			\left[ \nachoo{P_1} (1),\nachoo{P_2} (2)\right] = \left[ \nachoo{m_1 n_1} (1), \nachoo{m_2 n_2} (2) \right] = 2 i \eta_{m_1 n_1 m_2 n_2 r} \partial^r_1 \delta(1-2).
		\end{align*}
		Again, $SL(5)$ invariant tensors are proportional to 5 dimensional Levi-Civita tensors or their combinations. Since $\nachoo{P} = \frac{1}{2}\nachoo{[mn]}$ and $\eta_{P_1 P_2 r}$ should be symmetric in $P_1$ and $P_2$, it can be chosen to be
		\begin{align*}
			\eta_{P_1 P_2 r} = \eta_{m_1 n_1 m_2 n_2 r} = \epsilon_{m_1 n_1 m_2 n_2 r}.
		\end{align*}
		By construction,
		\begin{align*}
			& \left\lbrace \nachoo{D} , \nachoo{\Omega} \right\rbrace = \left\lbrace \nachoo{\alpha} , \nachoo{}^{\beta r} \right\rbrace = 2 i \tensor{\delta}{_{\alpha}^{\beta}} \partial^r_1 \delta(1-2)\\
			\Rightarrow &\ \eta_{D \Omega q} = \tensor{\eta}{_{\alpha}^{\beta r}_{q}} = \tensor{\delta}{_{\alpha}^{\beta}} \delta_{q}^{r}.
		\end{align*}
		As explained in Section \ref{GC}, the above structure constants and metrics are numerically the same as before and after introducing $H$ group element $g$. All we have to put in is $\eta_{S \Sigma a}$, and the rest of the structure constants and the metrics can be found by using the equations in Appendix \ref{AppA}. By construction, $\nacho{S}$ transforms all the indices the same way as usual $Spin(3,2)$ indices.
		
		We first use equation \ref{DDP} to find the only one that doesn't involve $\Sigma$ that's left, $\f{DP}{\Omega}$:
		\begin{align*}
			0 & = \tensor{f}{_{D_1 D_2}^{P'}}\eta_{P' P a} + \tensor{f}{_{D_1 P}^{\Omega'}}\eta_{D_2 \Omega' a}\\
			& = \tensor{f}{_{\alpha_1 \alpha_2}^{c'd'}}\eta_{c'd' c d a} + \tensor{f}{_{\alpha_1 c d \beta b}}\tensor{\eta}{_{\alpha_2} ^{\beta b} _{a}}\\
			\Rightarrow &\ \f{D P}{\Omega} = f_{\alpha c d \beta a} = \left( \gamma^{ef} \right)_{\alpha \beta} \epsilon_{e f c d a}.
		\end{align*}
		We now determine what $\eta_{S \Sigma a}$ is. As explained in Section \ref{GC}, $\eta_{S\Sigma a} = \tensor{\eta}{_{S}^{\widetilde{S} b}_{a}} = \tensor{\eta}{_{S}^{\widetilde{S}}}\delta_a^b$. In the 5-brane case, $\nacho{S}$ is $Spin(3,2)$ generators, hence {\bf S} is antisymmetric in its two indices. Using this property, we can conclude $\tensor{\eta}{_S ^{\widetilde{S}}} = \frac{1}{2}\delta_{cd} ^{ef}$. Using equation (\ref{fssigma}) we found the rest of two unsolved structure constants:
		\begin{align*}
			& \f{P_1 P_2}{\Sigma}\eta_{\Sigma S a} = \frac{1}{2}\f{S[P_1|}{P'}\eta_{P'|P_2]a}\\
			\Rightarrow &\ \f{P_1 P_2}{\Sigma} = f_{c_1 d_1 c_2 d_2 e f a} = - \left(\mathring{\eta}_{[c_1|[e}\epsilon_{f]|d_1]c_2 d_2 a} - \mathring{\eta}_{[c_2|[e}\epsilon_{f]|d_2]c_1 d_1 a}\right),\\
			& \f{D \Omega}{\Sigma}\eta_{\Sigma S a} = \frac{1}{2}\f{SD}{D'}\eta_{D' \Omega a} + \frac{1}{2}\f{S\Omega}{\Omega'}\eta_{\Omega' D a}\\
			\Rightarrow &\  \f{D \Omega}{\Sigma} = \tensor{f}{_{\alpha}^{\beta b}_{e f a}} = \frac{1}{2} \left( \gamma_{ef} \right)_{\alpha}^{~~\beta}\delta_{a}^{b}.
		\end{align*}
		The $\mathring{\eta}$ above is $SO(3,2)$ metric.
		
		The full commutation relations are listed in Appendix \ref{5braneCR}.
	\section{Conclusion}
	
		The method presented in this paper gives a consistent mathematical structure for higher dimensional brane current algebra (higher than 1) by construction, as opposed to the usual Jacobi identity method used in string theory \cite{Polacek:2013nla,Polacek:2014cva}. The main reason is that for 1-brane or string, on the ``metric'' the additional worldvolume index can take only one value, which is inert under $H$ transformation. For (higher dimensional) brane current algebra, the additional worldvolume indices can have more than one choice. However, in {\bf F}-theory worldvolume indices are also spacetime indices, therefore they all have to react to $H$-group transformations the same way spacetime indices do. This property makes the original construction unsuitable for the higher dimensional brane algebra (mentioned in the beginning of Section \ref{GC}). The method presented in this paper can be used in any finite dimensional brane. We've worked out the 5-brane case in detail.
	
		The method used in this paper is not just interesting by itself but also can be utilized for the following subjects:
		\begin{enumerate}[i)]
			\item Generalize the method to curved spacetime ({\bf F}-gravity).
			\item Analyze massive modes.
			\item Understand string field theory.
		\end{enumerate}
	
	\section*{Acknowledgments}
	
		CYJ thanks William D Linch III for helpful discussions. This work was supported in part by National Science Foundation Grant No. PHY-1316617.
	
	\appendix
	\section{Relating \texorpdfstring{$f$}{f}'s and \texorpdfstring{$\eta$}{\texteta}'s Using Jacobi \label{AppA}}
	
		The following are the complete list of all the relations between $f$'s and $\eta$'s. The ``zero modes'' means no derivative on delta function ones ($\delta^2$) and the ``oscillating modes'' means the ones that have a derivative on a delta function ($\delta \partial \delta$). Equations (\ref{zeroSSS} $\sim$ \ref{zeroSOmegaD}) show that $f$'s are invariant under group $H$, and equation (\ref{zeroDDOmega}) gives nothing new.
\vskip.2in
\noindent{\bf Zero modes:}
		\begin{align}
			& 0 = \tensor{f}{_{S_1 S_2}^{S'}}\tensor{f}{_{S' S_3}^{S _{4}}} + \tensor{f}{_{S_2 S_3}^{S'}}\tensor{f}{_{S' S_1}^{S _{4}}} + \tensor{f}{_{S_1 S_3}^{S'}}\tensor{f}{_{S_2 S'}^{S _{4}}}, \label{zeroSSS} \\
			& 0 = \tensor{f}{_{S_1 S_2}^{S'}}\tensor{f}{_{S' D_1}^{D_2}} + \tensor{f}{_{S_2 D_1}^{D'}}\tensor{f}{_{D' S_1}^{D_2}} + \tensor{f}{_{S_1 D_1}^{D'}}\tensor{f}{_{S_2 D'}^{D_2}},\\
			& 0 = \tensor{f}{_{S_1 S_2}^{S'}}\tensor{f}{_{S' P_1}^{P_2}} + \tensor{f}{_{S_2 P_1}^{P'}}\tensor{f}{_{P' S_1}^{P_2}} + \tensor{f}{_{S_1 P_1}^{P'}}\tensor{f}{_{S_2 P'}^{P_2}},\\
			& 0 = \tensor{f}{_{S_1 S_2}^{S'}}\tensor{f}{_{S' \Omega_1}^{\Omega_2}} + \tensor{f}{_{S_2 \Omega_1}^{\Omega'}}\tensor{f}{_{\Omega' S_1}^{\Omega_2}} + \tensor{f}{_{S_1 \Omega_1}^{\Omega'}}\tensor{f}{_{S_2 \Omega'}^{\Omega_2}},\\
			& 0 = \tensor{f}{_{S_1 S_2}^{S'}}\tensor{f}{_{S' \Sigma_1}^{\Sigma_2}} + \tensor{f}{_{S_2 \Sigma_1}^{\Sigma'}}\tensor{f}{_{\Sigma' S_1}^{\Sigma_2}} + \tensor{f}{_{S_1 \Sigma_1}^{\Sigma'}}\tensor{f}{_{S_2 \Sigma'}^{\Sigma_2}},\\
			& 0 = \tensor{f}{_{S D_1}^{D'}}\tensor{f}{_{D' D_2}^{P}} + \tensor{f}{_{D_1 D_2}^{P'}}\tensor{f}{_{P' S}^{P}} + \tensor{f}{_{S D_2}^{D'}}\tensor{f}{_{D_1 D'}^{P}},\\
			& 0 = \tensor{f}{_{S D}^{D'}}\tensor{f}{_{D' P}^{\Omega}} + \tensor{f}{_{D P}^{\Omega'}}\tensor{f}{_{\Omega' S}^{\Omega}} + \tensor{f}{_{S P}^{P'}}\tensor{f}{_{D P'}^{\Omega}},\\
			& 0 = \tensor{f}{_{S D}^{D'}}\tensor{f}{_{D' \Omega}^{\Sigma}} + \tensor{f}{_{D \Omega}^{\Sigma'}}\tensor{f}{_{\Sigma' S}^{\Sigma}} + \tensor{f}{_{S \Omega}^{\Omega'}}\tensor{f}{_{D \Omega'}^{\Sigma}},\\
			& 0 = \tensor{f}{_{S P}^{P'}}\tensor{f}{_{P' D}^{\Omega}} + \tensor{f}{_{P D}^{\Omega'}}\tensor{f}{_{\Omega' S}^{\Omega}} + \tensor{f}{_{S D}^{D'}}\tensor{f}{_{P D'}^{\Omega}},\\
			& 0 = \tensor{f}{_{S P_1}^{P'}}\tensor{f}{_{P' P_2}^{\Sigma}} + \tensor{f}{_{P_1 P_2}^{\Sigma'}}\tensor{f}{_{\Sigma' S}^{\Sigma}} + \tensor{f}{_{S P_2}^{P'}}\tensor{f}{_{P_1 P'}^{\Sigma}},\\
			& 0 = \tensor{f}{_{S \Omega}^{\Omega'}}\tensor{f}{_{\Omega' D}^{\Sigma}} + \tensor{f}{_{\Omega D}^{\Sigma'}}\tensor{f}{_{\Sigma' S}^{\Sigma}} + \tensor{f}{_{S D}^{D'}}\tensor{f}{_{\Omega D'}^{\Sigma}},\label{zeroSOmegaD}\\
			& 0 = \tensor{f}{_{D_1 D_2}^{P'}}\tensor{f}{_{P' D_3}^{\Omega}} + \tensor{f}{_{D_2 D_3}^{P'}}\tensor{f}{_{P' D_1}^{\Omega}} + \tensor{f}{_{D_3 D_1}^{P'}}\tensor{f}{_{P' D_2}^{\Omega}},\label{zeroDDOmega}\\
			& 0 = \tensor{f}{_{D_1 D_2}^{P'}}\tensor{f}{_{P' P}^{\Sigma}} - \tensor{f}{_{D_2 P}^{\Omega'}}\tensor{f}{_{\Omega' D_1}^{\Sigma}} + \tensor{f}{_{P D_1}^{\Omega'}}\tensor{f}{_{\Omega' D_2}^{\Sigma}}.
		\end{align}
		
\noindent{\bf Oscillating modes: }
		\begin{align}
			& 0 = \tensor{f}{_{S_1 S_2}^{S'}}\eta_{S' \Sigma a} + \tensor{f}{_{S_1 \Sigma}^{\Sigma'}}\eta_{S_2 \Sigma' a} + \tensor{f}{_{S_1 a}^{b}}\eta_{S_2 \Sigma b},\label{SSigma}\\
			& 0 = \tensor{f}{_{S D}^{D'}}\eta_{D' \Omega a} + \tensor{f}{_{S \Omega}^{\Omega'}}\eta_{D \Omega' a} + \tensor{f}{_{S a}^{b}}\eta_{D \Omega b},\label{SDOmega}\\
			& 0 = \tensor{f}{_{S P_1}^{P'}}\eta_{P' P_2 a} + \tensor{f}{_{S P_2}^{P'}}\eta_{P_1 P' a} + \tensor{f}{_{S a}^{b}}\eta_{P_1 P_2 b},\label{SPP}\\
			& 0 = \tensor{f}{_{D S}^{D'}}\eta_{D' \Omega a} + \tensor{f}{_{D \Omega}^{\Sigma'}}\eta_{S \Sigma' a} - \frac{1}{2}\tensor{f}{_{S a}^{b}}\eta_{D \Omega b},\label{DSOmega}\\
			& 0 = \tensor{f}{_{D_1 D_2}^{P'}}\eta_{P' P a} + \tensor{f}{_{D_1 P}^{\Omega'}}\eta_{D_2 \Omega' a},\label{DDP}\\
			& 0 = \tensor{f}{_{P_1 S}^{P'}}\eta_{P' P_2 a} + \tensor{f}{_{P_1 P_2}^{\Sigma'}}\eta_{S \Sigma' a} - \frac{1}{2}\tensor{f}{_{S a}^{b}}\eta_{P_1 P_2 b},\label{PSP}\\
			& 0 = \tensor{f}{_{P D_1}^{\Omega'}}\eta_{\Omega' D_2 a} + \tensor{f}{_{P D_2}^{\Omega'}}\eta_{D_1 \Omega' a},\\
			& 0 = \tensor{f}{_{\Omega S}^{\Omega'}}\eta_{\Omega' D a} + \tensor{f}{_{\Omega D}^{\Sigma'}}\eta_{S \Sigma' a} - \frac{1}{2}\tensor{f}{_{S a}^{b}}\eta_{\Omega D b},\label{OmegaSD}\\
			& 0 = \tensor{f}{_{\Sigma S_1}^{\Sigma'}}\eta_{\Sigma' S_2 a} + \tensor{f}{_{\Sigma S_2}^{\Sigma'}}\eta_{S_1 \Sigma' a}.
		\end{align}
		
	\section{5-Brane Commutation Relations\label{5braneCR}}

		This appendix shows all the nonvanishing commutation relations for the 5-brane.
{\openup5pt
		\begin{enumerate}
			\item $ \displaystyle \Big\lbrace \nacho{D_1}(1) , \nacho{D_2}(2) \Big\rbrace = \Big\lbrace \nacho{\alpha_1}(1) , \nacho{\alpha_2}(2) \Big\rbrace$\\
			$ = i \left(\gamma^{mn}\right)_{\alpha_1 \alpha_2} \nachoo{mn} \delta(1-2) = i \f{D_1 D_2}{P}\nacho{P} \delta(1-2).$
			\item $ \displaystyle \Big[ \nacho{D}(1) , \nacho{P}(2) \Big] = \Big[ \nacho{\alpha}(1) , \nacho{cd}(2) \Big]$\\
			$ = i \left( \gamma^{ef} \right)_{\alpha \beta} \epsilon_{e f c d a} \uppernacho {\beta a} \delta(1-2)= i \f{D_1 P}{\Omega}\nacho{\Omega}\delta(1-2).$
			\item $ \displaystyle \Big[ \nacho{P_1}(1) , \nacho{P_2}(2) \Big] = \Big[ \nacho{c_1 d_1}(1) , \nacho{c_2 d_2}(2) \Big]$\\
			$ = i \left( - \mathring{\eta}_{[c_1|[e}\epsilon_{f]|d_1]c_2 d_2 a} + \mathring{\eta}_{[c_2|[e}\epsilon_{f]|d_2]c_1 d_1 a}\right) \uppernacho{efa} \delta(1-2) $\\
			\phantom{m}$ + i \epsilon_{c_1 d_1 c_2 d_2 a} \uppernacho{a} ((1) - (2) \delta(1-2) $\\
			$ = i \f{P_1 P_2}{\Sigma}\nacho{\Sigma} \delta(1-2) + i \eta_{P_1 P_2 a} \uppernacho{a} ((1) - (2)) \delta(1-2).$
			\item $ \displaystyle \Big\lbrace \nacho{D}(1) , \nacho{\Omega}(2) \Big\rbrace = \Big\lbrace \nacho{\alpha}(1) , \uppernacho{\beta b}(2) \Big\rbrace$\\
			$ = \frac{i}{2} \left( \gamma_{ef} \right)_{\alpha}^{~~\beta} \uppernacho{efb} \delta(1-2) + i \tensor{\delta}{_{\alpha}^{\beta}} \delta_{a}^{b} \uppernacho{a} ((1) - (2)) \delta(1-2)$\\
			$ = i \f{D \Omega}{\Sigma}\nacho{\Sigma} \delta(1-2) + i \eta_{D \Omega a} \uppernacho{a} ((1) - (2)) \delta(1-2).$
			\item $ \displaystyle \Big[ \nacho{S}(1) , \nacho{D}(2) \Big]  = \Big[ \nacho{ef}(1) , \nacho{\alpha}(2) \Big]$\\
			$ = \frac{i}{4} \left( \gamma_{ef} \right)_{\alpha}^{~~\rho} \nacho{\rho} \delta(1-2) = i \f{S D}{D'}\nacho{D'} \delta(1-2).$
			\item $ \displaystyle \Big[ \nacho{S}(1) , \nacho{P}(2) \Big] = \Big[ \nacho{ef}(1) , \nacho{cd}(2) \Big]$\\
			$ = - i \mathring{\eta}_{[c|[e} \delta_{f]|d]}^{c'd'}\nacho{c'd'} \delta(1-2) = i \f{S P}{P'}\nacho{P'} \delta(1-2).$
			\item $ \displaystyle \Big[ \nacho{S}(1) , \nacho{\Omega}(2) \Big] = \Big[ \nacho{ef}(1) , \uppernacho{\beta b}(2) \Big]$\\
			$ = i\left[ -\frac{1}{4} \left( \gamma_{ef} \right)_{\rho}^{~~\beta} \delta_{a}^{b} + \delta_{\rho}^{~\beta}\delta_{[e}^{b}\mathring{\eta}_{f]a} \right]\uppernacho{\rho a} \delta(1-2)= i \f{S \Omega}{\Omega'}\nacho{\Omega'} \delta(1-2).$
			\item $ \displaystyle \Big[ \nacho{S}(1) , \nacho{S}(2) \Big] = \Big[ \nacho{e_1 f_1}(1) , \nacho{e_2 f_2}(2) \Big]$\\
			$ = - i \mathring{\eta}_{[e_2|[e_1} \delta_{f_1]|f_2]}^{e'f'}\nacho{e'f'} \delta(1-2) = i \f{S S}{S'}\nacho{S'} \delta(1-2).$
			\item $ \displaystyle \Big[ \nacho{S}(1) , \nacho{\Sigma}(2) \Big] = \Big[ \nacho{ef}(1) , \uppernacho{gha}(2) \Big]$\\
			$ = i \mathring{\eta}_{[g'|[e} \delta_{f]|h']}^{gh}\delta_{b}^{a}\uppernacho{g'h'b} \delta(1-2) + i \mathring{\eta}_{b'[e}\delta_{f]}^{a}\uppernacho{ghb'} \delta(1-2) + 2 i \delta_{ef}^{gh}\uppernacho{a} (2) \delta(1-2)$\\
			$ = i \f{S \Sigma}{\Sigma'}\nacho{\Sigma'} \delta(1-2)+ 2 i \eta_{S \Sigma a} \uppernacho{a} (2) \delta(1-2).$
		\end{enumerate}
}

	\bibliographystyle{utphys}
	\bibliography{references}

\providecommand{\href}[2]{#2}\begingroup\raggedright\begin{thebibliography}{10}

\bibitem{Giveon:1994fu}
A.~Giveon, M.~Porrati, and E.~Rabinovici, ``{Target space duality in string
  theory},'' \href{http://dx.doi.org/10.1016/0370-1573(94)90070-1}{{\em Phys.
  Rept.} {\bfseries 244} (1994) 77--202},
\href{http://arxiv.org/abs/hep-th/9401139}{{\ttfamily arXiv:hep-th/9401139
  [hep-th]}}.

\bibitem{Seiberg:1994pq}
N.~Seiberg, ``{Electric - magnetic duality in supersymmetric nonAbelian gauge
  theories},'' \href{http://dx.doi.org/10.1016/0550-3213(94)00023-8}{{\em Nucl.
  Phys.} {\bfseries B435} (1995) 129--146},
\href{http://arxiv.org/abs/hep-th/9411149}{{\ttfamily arXiv:hep-th/9411149
  [hep-th]}}.

\bibitem{Hull:1994ys}
C.~M. Hull and P.~K. Townsend, ``{Unity of superstring dualities},''
  \href{http://dx.doi.org/10.1016/0550-3213(94)00559-W}{{\em Nucl. Phys.}
  {\bfseries B438} (1995) 109--137},
\href{http://arxiv.org/abs/hep-th/9410167}{{\ttfamily arXiv:hep-th/9410167
  [hep-th]}}.

\bibitem{Witten:1995ex}
E.~Witten, ``{String theory dynamics in various dimensions},''
  \href{http://dx.doi.org/10.1016/0550-3213(95)00158-O}{{\em Nucl. Phys.}
  {\bfseries B443} (1995) 85--126},
\href{http://arxiv.org/abs/hep-th/9503124}{{\ttfamily arXiv:hep-th/9503124
  [hep-th]}}.

\bibitem{Horava:1995qa}
P.~Horava and E.~Witten, ``{Heterotic and type I string dynamics from
  eleven-dimensions},''
  \href{http://dx.doi.org/10.1016/0550-3213(95)00621-4}{{\em Nucl. Phys.}
  {\bfseries B460} (1996) 506--524},
\href{http://arxiv.org/abs/hep-th/9510209}{{\ttfamily arXiv:hep-th/9510209
  [hep-th]}}.

\bibitem{Cremmer:1997ct}
E.~Cremmer, B.~Julia, H.~Lu, and C.~N. Pope, ``{Dualization of dualities.
  1.},'' \href{http://dx.doi.org/10.1016/S0550-3213(98)00136-9}{{\em Nucl.
  Phys.} {\bfseries B523} (1998) 73--144},
\href{http://arxiv.org/abs/hep-th/9710119}{{\ttfamily arXiv:hep-th/9710119
  [hep-th]}}.

\bibitem{Obers:1998fb}
N.~A. Obers and B.~Pioline, ``{U duality and M theory},''
  \href{http://dx.doi.org/10.1016/S0370-1573(99)00004-6}{{\em Phys. Rept.}
  {\bfseries 318} (1999) 113--225},
\href{http://arxiv.org/abs/hep-th/9809039}{{\ttfamily arXiv:hep-th/9809039
  [hep-th]}}.

\bibitem{Vafa:1996xn}
C.~Vafa, ``{Evidence for F theory},''
  \href{http://dx.doi.org/10.1016/0550-3213(96)00172-1}{{\em Nucl. Phys.}
  {\bfseries B469} (1996) 403--418},
\href{http://arxiv.org/abs/hep-th/9602022}{{\ttfamily arXiv:hep-th/9602022
  [hep-th]}}.

\bibitem{Siegel:1993xq}
W.~Siegel, ``{Two vierbein formalism for string inspired axionic gravity},''
  \href{http://dx.doi.org/10.1103/PhysRevD.47.5453}{{\em Phys. Rev.} {\bfseries
  D47} (1993) 5453--5459},
\href{http://arxiv.org/abs/hep-th/9302036}{{\ttfamily arXiv:hep-th/9302036
  [hep-th]}}.

\bibitem{Siegel:1993th}
W.~Siegel, ``{Superspace duality in low-energy superstrings},''
  \href{http://dx.doi.org/10.1103/PhysRevD.48.2826}{{\em Phys. Rev.} {\bfseries
  D48} (1993) 2826--2837},
\href{http://arxiv.org/abs/hep-th/9305073}{{\ttfamily arXiv:hep-th/9305073
  [hep-th]}}.

\bibitem{Siegel:1993bj}
W.~Siegel, ``{Manifest duality in low-energy superstrings},'' in {\em {In
  *Berkeley 1993, Proceedings, Strings '93* 353-363, and State U. New York
  Stony Brook - ITP-SB-93-050 (93,rec.Sep.) 11 p. (315661)}}.
\newblock 1993.
\newblock
\href{http://arxiv.org/abs/hep-th/9308133}{{\ttfamily arXiv:hep-th/9308133
  [hep-th]}}.
\newblock

\bibitem{Hull:2009mi}
C.~Hull and B.~Zwiebach, ``{Double Field Theory},''
  \href{http://dx.doi.org/10.1088/1126-6708/2009/09/099}{{\em JHEP} {\bfseries
  09} (2009) 099},
\href{http://arxiv.org/abs/0904.4664}{{\ttfamily arXiv:0904.4664 [hep-th]}}.

\bibitem{Hatsuda:2012uk}
M.~Hatsuda and T.~Kimura, ``{Canonical approach to Courant brackets for
  D-branes},'' \href{http://dx.doi.org/10.1007/JHEP06(2012)034}{{\em JHEP}
  {\bfseries 06} (2012) 034},
\href{http://arxiv.org/abs/1203.5499}{{\ttfamily arXiv:1203.5499 [hep-th]}}.

\bibitem{Hatsuda:2013dya}
M.~Hatsuda and K.~Kamimura, ``{M5 algebra and SO(5,5) duality},''
  \href{http://dx.doi.org/10.1007/JHEP06(2013)095}{{\em JHEP} {\bfseries 06}
  (2013) 095},
\href{http://arxiv.org/abs/1305.2258}{{\ttfamily arXiv:1305.2258 [hep-th]}}.

\bibitem{Linch:2015lwa}
W.~D. Linch and W.~Siegel, ``{F-theory Superspace},''
\href{http://arxiv.org/abs/1501.02761}{{\ttfamily arXiv:1501.02761 [hep-th]}}.

\bibitem{Linch:2015fya}
W.~D. Linch, III and W.~Siegel, ``{F-theory from Fundamental Five-branes},''
\href{http://arxiv.org/abs/1502.00510}{{\ttfamily arXiv:1502.00510 [hep-th]}}.

\bibitem{Linch:2015fca}
W.~D. Linch and W.~Siegel, ``{Critical Super F-theories},''
\href{http://arxiv.org/abs/1507.01669}{{\ttfamily arXiv:1507.01669 [hep-th]}}.

\bibitem{Linch:2015qva}
W.~D. Linch and W.~Siegel, ``{F-theory with Worldvolume Sectioning},''
\href{http://arxiv.org/abs/1503.00940}{{\ttfamily arXiv:1503.00940 [hep-th]}}.

\bibitem{Park:2013gaj}
J.-H. Park and Y.~Suh, ``{U-geometry: SL(5)},''
  \href{http://dx.doi.org/10.1007/JHEP11(2013)210,
  10.1007/JHEP04(2013)147}{{\em JHEP} {\bfseries 04} (2013) 147},
  \href{http://arxiv.org/abs/1302.1652}{{\ttfamily arXiv:1302.1652 [hep-th]}}.
[Erratum: JHEP11,210(2013)].

\bibitem{Polacek:2013nla}
M.~Pol\'{a}\v{c}ek and W.~Siegel, ``{Natural curvature for manifest
  T-duality},'' \href{http://dx.doi.org/10.1007/JHEP01(2014)026}{{\em JHEP}
  {\bfseries 01} (2014) 026},
\href{http://arxiv.org/abs/1308.6350}{{\ttfamily arXiv:1308.6350 [hep-th]}}.

\bibitem{Polacek:2014cva}
M.~Pol\'{a}\v{c}ek and W.~Siegel, ``{T-duality off shell in 3D Type II
  superspace},'' \href{http://dx.doi.org/10.1007/JHEP06(2014)107}{{\em JHEP}
  {\bfseries 06} (2014) 107},
\href{http://arxiv.org/abs/1403.6904}{{\ttfamily arXiv:1403.6904 [hep-th]}}.

\end{thebibliography}\endgroup
\end{document}